\documentclass[acmsmall,nonacm]{acmart}

\usepackage{booktabs}
\usepackage{tabularx}
\usepackage{array}
\usepackage{tikz}
\usepackage{enumitem}
\usetikzlibrary{arrows.meta,positioning,fit}

\setcopyright{rightsretained}
\makeatletter
\def\@acmBadgeL@image{}
\def\@acmBadgeR@image{}
\makeatother

\newcolumntype{Y}{>{\raggedright\arraybackslash}X}
\setlist[itemize]{leftmargin=1.2em, itemsep=0.2em, topsep=0.25em}

\usepackage{xcolor}
\usetikzlibrary{arrows.meta,positioning,calc,shapes.geometric,fit,backgrounds}

\definecolor{dtblue}{RGB}{31,95,158}
\definecolor{dtgreen}{RGB}{34,134,90}
\definecolor{dtamber}{RGB}{196,128,20}
\definecolor{dtgray}{RGB}{90,98,110}
\definecolor{dtlight}{RGB}{236,242,248}
\definecolor{dtlightg}{RGB}{233,245,238}

\title{Digital Twins Need Feedback}

\author{Guo-Qiang Zhang}
\email{guo-qiang.zhang@uth.tmc.edu}
\affiliation{%
  \institution{The University of Texas Health Science Center at Houston}
  \city{Houston}
  \state{Texas}
  \country{USA}
}

\begin{document}

\begin{abstract}
Digital twins are too often described as realistic simulations, anatomical avatars, dashboards, or data mirrors. Those artifacts can be useful, but they miss the defining property of a digital twin: bidirectional feedback between a physical counterpart and a virtual counterpart. The physical system continuously updates the virtual one; the virtual system informs actions that change measurement, intervention, operation, or governance in the physical world. We propose such a  bidirectional feedback as the organizing principle for digital twins and apply  it to a nested, multi-scale hierarchy of biological and social organization, in which lower-level units combine into higher-level systems, producing desirable properties at each level, from cells and tissues to organs, individuals, organizations, and population at large.
Neuroinformatics is a stress test for this view because brain health, dementia, epilepsy, and other neurological diseases require the integration of cells, circuits, behavior, care pathways, and the translation of discovery to practice. 
Examples from epilepsy care and consortium-scale brain-cell atlas production show that digital twinning is not merely multi-scale modeling. 
It is a rich, multidisciplinary paradigm of computing for designing, governing, and driving feedback loops that turn data into accountable action.
\end{abstract}

\keywords{digital twins, bidirectional feedback, neuroinformatics, learning health systems, cyber-physical systems, real-world data, biomedical informatics}

\begin{CCSXML}
<ccs2012>
   <concept>
       <concept_id>10010405.10010476.10010477</concept_id>
       <concept_desc>Applied computing~Health informatics</concept_desc>
       <concept_significance>500</concept_significance>
   </concept>
   <concept>
       <concept_id>10010520.10010553.10010562</concept_id>
       <concept_desc>Computer systems organization~Embedded and cyber-physical systems</concept_desc>
       <concept_significance>300</concept_significance>
   </concept>
   <concept>
       <concept_id>10002951.10003260.10003261</concept_id>
       <concept_desc>Information systems~Data management systems</concept_desc>
       <concept_significance>300</concept_significance>
   </concept>
   <concept>
       <concept_id>10010147.10010257</concept_id>
       <concept_desc>Computing methodologies~Machine learning</concept_desc>
       <concept_significance>100</concept_significance>
   </concept>
</ccs2012>
\end{CCSXML}

\ccsdesc[500]{Applied computing~Health informatics}
\ccsdesc[300]{Computer systems organization~Embedded and cyber-physical systems}
\ccsdesc[300]{Information systems~Data management systems}
\ccsdesc[100]{Computing methodologies~Machine learning}

\maketitle

\section{A cockpit is not enough}

The traditional concept of a digital twin is straightforward to explain: a pilot sits before an instrument panel that represents the state of an aircraft. The panel compresses a complex physical machine into actionable signals. Airspeed, altitude, attitude, fuel, weather, terrain, and engine state are arranged so 
the pilot can make timely decisions.

The modern digital twin paradigm emerged largely from manufacturing and aerospace, where virtual replicas of complex physical assets were developed to support design, simulation, operational monitoring, proactive management, and lifecycle optimization~\cite{Grieves2014DigitalTwin,Glaessgen2012NASA,Fuller2020DigitalTwin}. The image of the aircraft instrument panel 
 is useful, but incomplete. A display in front of a pilot is not yet a digital twin. It becomes twin-like when the aircraft and its representation are coupled in a closed loop. Sensor measurements update the representation; the representation informs pilot action, autopilot control, maintenance decisions, flight-path planning, or the next measurements to collect. Feedback is the difference between an instrument and the instrumented system: an instrument only observes; an instrumented system observes and responds through a feedback loop.

This distinction matters because the term ``digital twin'' is now used for almost anything computationally related to a physical object: a simulation, a dashboard, a three-dimensional visualization,  a predictive model, a control algorithm, or a digital artifact~\cite{Mao}. The National Academies' report on digital twins provides a general framework: a digital twin couples computational models with a physical counterpart in a system which is dynamically updated through bidirectional information flows as conditions change~\cite{NASEM2024DigitalTwins}. This feedback requirement separates digital twins from ordinary modeling.

Computing already has related, but not identical concepts. Cyber-physical systems emphasize the integration of computation, networking, and physical processes~\cite{LeeSeshia2017CPS}. The Internet of Things emphasizes networked sensors, devices, and services~\cite{Atzori2010IoT}. Simulation has long supported engineering, design, and training. Digital twins borrow from all three, but their novelty lies in making the virtual counterpart an operational participant in a feedback relationship with a specific physical counterpart. The physical entity is not merely a data source; the virtual counterpart is not merely an observer.

\begin{quote}
\textcolor{blue}{\bf Bidirectional feedback.} 
\textcolor{teal}{Let $S_t$ be the state of a physical entity at time $t$ and $\hat S_t$ the
state of its virtual counterpart. \emph{Feed-forward}:
$\hat S_t = U(\hat S_{t-1}, o_t)$, where $o_t$ is an observation drawn
from $S_t$ and $U$ is an update operator. \emph{Feed-backward}:
$a_t = D(\hat S_t)$, where $D$ maps the twin's state to a decision or
control $a_t$ that influences $S_{t+1}$. A static avatar fixes
$\hat S$; a shadow implements $U$ but not $D$; an open-loop protocol
implements $D$ but not $U$. A digital twin $F$ combines both $D$ and $U$, so that the
composite operator $S_{t+1}=F(S_t, D(U(\hat S_{t-1},o_t)))$ is genuinely
closed. Everything that matters to the twining, such as decision, response, latency, effect, is a property of this composition, not of $U$ or $D$ alone.}
\end{quote}

The health domain exposes the cost of blurring these distinctions; recent medical reviews emphasize similar distinctions between models, data streams, and twinned clinical systems~\cite{Laubenbacher2024Medicine,Zhang2024HealthcareDT}. A hospital dashboard that shows the number of daily encounters  is not a hospital digital twin. A brain atlas is not automatically a brain digital twin. A multiscale systems-biology model is not automatically a medical digital twin. Each can become a component of a digital twin only when it is embedded in a reliable, continuously evolving loop: sensed state, virtual representation, prediction or inference, action, measurement of consequence, and update. 

\section{Feedback is the first-class object}

Digital twins should be evaluated less by how realistic they look, but more by what feedback they support. 
A useful feedback description answers five questions.

First, what is the physical counterpart? It may be an organ, a patient, a care team, a hospital workflow, a data-generating consortium, or a population. Second, what state of that counterpart is represented digitally, and at what semantic resolution? Third, how does the physical state update the digital state? Fourth, how does the digital state influence the physical system? Fifth, how are errors, uncertainty, accountability, consent, and responsibility handled when the loop acts?

These questions reveal why digital twins differ from standard multi-scale systems biology. Systems biology has built powerful models that connect molecular, cellular, tissue, organ, and organism scales. A digital twin must add actionability and coupling. It must define which model outputs matter to a physical counterpart, how they affect measurement or intervention, who is authorized to act, and how the outcome updates the model. A twin is not just a multi-scale model of a system. It is a governed feedback system for a coupled system: one part is physical, and the other virtual. 

They also reveal why digital twins differ from conventional cyber-physical systems. A pacemaker, ventilator, or robot is already cyber-physical. A digital twin adds a persistent, semantically rich virtual counterpart that can accumulate history, compare possible futures, support human decisions, explain its own evidence, and connect local control to enterprise learning. A twin should therefore be treated as a systems artifact, a data artifact, a model artifact, and an institutional artifact at the same time.

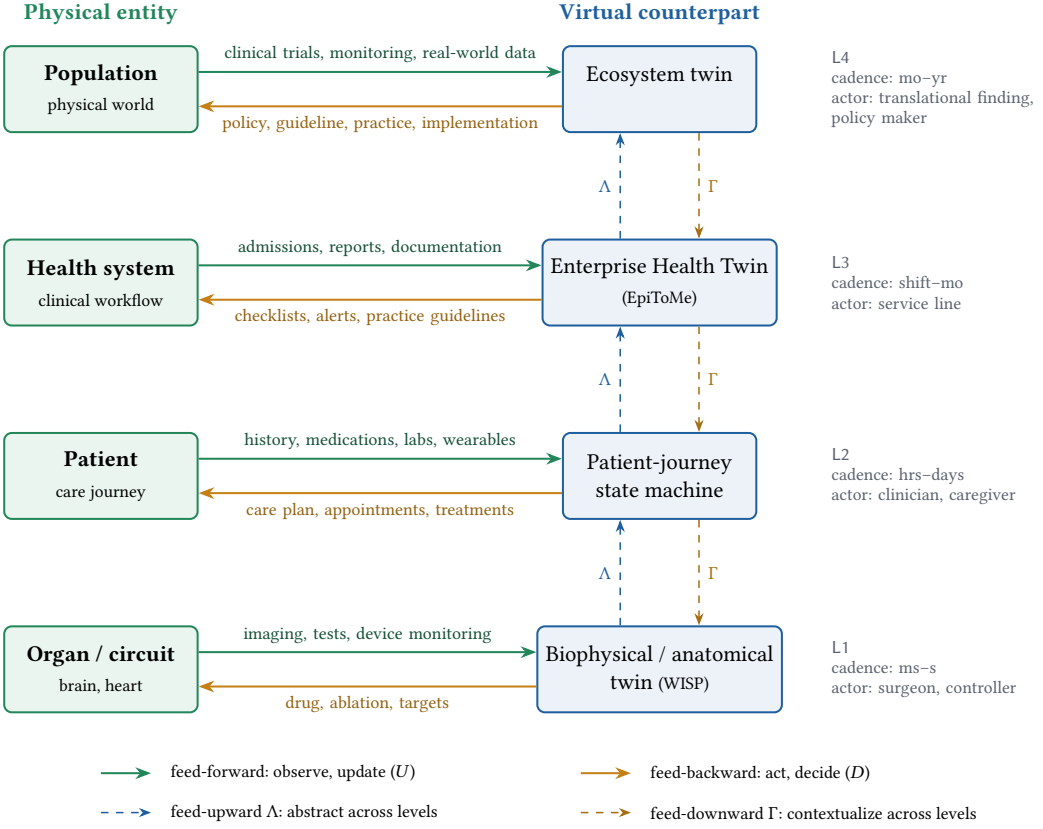
\begin{figure*}[t]
\centering
\resizebox{\textwidth}{!}{%
\begin{tikzpicture}[
  font=\small,
  phys/.style ={rounded corners=3pt, draw, thick, minimum width=2.7cm,
                minimum height=1.2cm, align=center, fill=dtlightg, draw=dtgreen, text=black},
  twin/.style ={rounded corners=3pt, draw, thick, minimum width=2.7cm,
                minimum height=1.2cm, align=center, fill=dtlight, draw=dtblue, text=black},
  ff/.style={-{Stealth[length=2.2mm]}, draw=dtgreen, thick},
  fb/.style={-{Stealth[length=2.2mm]}, draw=dtamber, thick},
  fu/.style={-{Stealth[length=1.7mm]}, draw=dtblue, dashed, semithick},
  fd/.style={-{Stealth[length=1.7mm]}, draw=dtamber!85!black, dashed, semithick},
  flab/.style={font=\scriptsize, text=dtgreen!60!black, align=center},
  blab/.style={font=\scriptsize, text=dtamber!75!black, align=center},
  ann/.style={font=\scriptsize, text=dtgray, align=left, text width=3.0cm},
  vlab/.style={font=\scriptsize\itshape}
]
\def\PX{0}
\def\VX{7.8}
% physical entities (left) and virtual twins (right); broadest level on top
\node[phys] (c)  at (\PX,0)    {{\textbf{Population}}\\[1pt]{\scriptsize physical world}};  %Population}\\ discovery-to-impact
\node[phys] (e)  at (\PX,-2.7) {{\textbf{Health system}}\\[1pt]{\scriptsize clinical workflow}};
\node[phys] (p)  at (\PX,-5.4) {{\textbf{Patient}}\\[1pt]{\scriptsize care journey}};
\node[phys] (o)  at (\PX,-8.1) {{\textbf{Organ / circuit}}\\[1pt]{\scriptsize brain, heart}};
\node[twin] (c2) at (\VX,0)    {Ecosystem twin\\[1pt]{\scriptsize }};
\node[twin] (e2) at (\VX,-2.7) {Enterprise Health Twin\\[1pt]{\scriptsize (EpiToMe)}};
\node[twin] (p2) at (\VX,-5.4) {Patient-journey\\state machine};
\node[twin] (o2) at (\VX,-8.1) {Biophysical / anatomical\\twin {\scriptsize (WISP)}};
\node[font=\bfseries\small, text=dtgreen] at (\PX,1.05) {Physical entity};
\node[font=\bfseries\small, text=dtblue]  at (\VX,1.05) {Virtual counterpart};
% within-level loops: feed-forward (green, U) + feed-backward (amber, D), each uniquely labeled
\draw[ff] ($(c.east)+(0,0.24)$)  -- node[flab,above,text width=4.7cm]{clinical trials, monitoring, real-world data} ($(c2.west)+(0,0.24)$);
\draw[fb] ($(c2.west)+(0,-0.24)$) -- node[blab,below,text width=4.7cm]{ policy, guideline, practice,  implementation} ($(c.east)+(0,-0.24)$);
\draw[ff] ($(e.east)+(0,0.24)$)  -- node[flab,above,text width=4.7cm]{admissions, reports, documentation} ($(e2.west)+(0,0.24)$);
\draw[fb] ($(e2.west)+(0,-0.24)$) -- node[blab,below,text width=4.7cm]{checklists, alerts, practice guidelines} ($(e.east)+(0,-0.24)$);
\draw[ff] ($(p.east)+(0,0.24)$)  -- node[flab,above,text width=4.7cm]{history, medications, labs, wearables} ($(p2.west)+(0,0.24)$);
\draw[fb] ($(p2.west)+(0,-0.24)$) -- node[blab,below,text width=4.7cm]{care plan, appointments, treatments} ($(p.east)+(0,-0.24)$);
\draw[ff] ($(o.east)+(0,0.24)$)  -- node[flab,above,text width=4.7cm]{imaging, tests, device monitoring} ($(o2.west)+(0,0.24)$);
\draw[fb] ($(o2.west)+(0,-0.24)$) -- node[blab,below,text width=4.7cm]{drug, ablation, targets} ($(o.east)+(0,-0.24)$);
% loop tags, cadence, acting constituency
\node[ann] at (11.7,0)    {$\mathsf{L4}$\\cadence: mo--yr\\actor: translational finding, policy maker};
\node[ann] at (11.7,-2.7) {$\mathsf{L3}$\\cadence: shift--mo\\actor: service line};
\node[ann] at (11.7,-5.4) {$\mathsf{L2}$\\cadence: hrs--days\\actor: clinician, caregiver};
\node[ann] at (11.7,-8.1) {$\mathsf{L1}$\\cadence: ms--s\\actor: surgeon, controller};
% across-level holarchic currents on the twin column: feed-upward (Lambda) up, feed-downward (Gamma) down
\draw[fu] ($(e2.north)+(-0.55,0)$) -- node[vlab,text=dtblue,left=0pt]{$\Lambda$} ($(c2.south)+(-0.55,0)$);
\draw[fd] ($(c2.south)+(0.55,0)$)  -- node[vlab,text=dtamber!85!black,right=0pt]{$\Gamma$} ($(e2.north)+(0.55,0)$);
\draw[fu] ($(p2.north)+(-0.55,0)$) -- node[vlab,text=dtblue,left=0pt]{$\Lambda$} ($(e2.south)+(-0.55,0)$);
\draw[fd] ($(e2.south)+(0.55,0)$)  -- node[vlab,text=dtamber!85!black,right=0pt]{$\Gamma$} ($(p2.north)+(0.55,0)$);
\draw[fu] ($(o2.north)+(-0.55,0)$) -- node[vlab,text=dtblue,left=0pt]{$\Lambda$} ($(p2.south)+(-0.55,0)$);
\draw[fd] ($(p2.south)+(0.55,0)$)  -- node[vlab,text=dtamber!85!black,right=0pt]{$\Gamma$} ($(o2.north)+(0.55,0)$);
% legend
\begin{scope}[shift={(0,-9.55)}]
  \draw[ff] (0,0) -- (0.7,0);
  \node[anchor=west, font=\scriptsize, text=black] at (0.85,0) {feed-forward: observe, update ($U$)};
  \draw[fb] (6.7,0) -- (7.4,0);
  \node[anchor=west, font=\scriptsize, text=black] at (7.55,0) {feed-backward: act, decide ($D$)};
  \draw[fu] (0,-0.55) -- (0.7,-0.55);
  \node[anchor=west, font=\scriptsize, text=black] at (0.85,-0.55) {feed-upward $\Lambda$: abstract across levels};
  \draw[fd] (6.7,-0.55) -- (7.4,-0.55);
  \node[anchor=west, font=\scriptsize, text=black] at (7.55,-0.55) {feed-downward $\Gamma$: contextualize across levels};
\end{scope}
\end{tikzpicture}%
}
\caption{\textbf{Twinning as a holarchy of bidirectional loops.} Each level couples a physical entity (green) to a virtual counterpart (blue) through a \emph{feed-forward} current (green; the update operator $U$, carrying observations into the twin) and a \emph{feed-backward} current (amber; the decision operator $D$, carrying actions back onto the world). The eight horizontal labels name the specific information exchanged at each level, so no two arrows mean the same thing. The four loops are tagged $\mathsf{L1}$ (organ/circuit), $\mathsf{L2}$ (patient), $\mathsf{L3}$ (health system), and $\mathsf{L4}$ (consortium/population), drawn broadest-on-top; what changes down the stack is each loop's \emph{cadence} and the \emph{constituency} that acts (right margin). The dashed currents on the twin column are the holarchic couplings: \emph{feed-upward} $\Lambda$ abstracts a lower twin's state into evidence for the level above (operational phenotypes $\rightarrow$ enterprise; structured data $\rightarrow$ consortium), while \emph{feed-downward} $\Gamma$ codifies higher-level knowledge into lower-level decision support (consortium atlases and common data elements $\rightarrow$ organ- and patient-level interpretation). The physical entities are themselves nested (organ $\subset$ patient $\subset$ health system $\subset$ population); the twins are coupled only through $\Lambda$ and $\Gamma$, which must preserve semantic resolution rather than manufacture precision. Exemplars: $\mathsf{L1}$, the Virtual Epileptic Patient and SEEG planning (WISP)~\cite{Jirsa2017VEP,Chou2025WISP}; $\mathsf{L3}$, EpiToMe~\cite{Tao2021EpiToMe,Zhang2026EHT}; $\mathsf{L4}$, Tissue-to-Bytes~\cite{Tao2025TissueToBytes}.}
\Description{A four-level hierarchy of digital-twin feedback loops with physical entities on the left and virtual twins on the right, broadest (consortium) on top and organ/circuit at the bottom. Each level has a green feed-forward arrow into the twin and an amber feed-backward arrow back to the world, each labeled with the specific information exchanged. Dashed arrows between adjacent twins show feed-upward abstraction and feed-downward contextualization across levels.}
\label{fig:hierarchy}
\end{figure*}

The most productive shift is to make feedback itself the design object. A \emph{feedback contract} should specify the state variables exchanged, the latency and reliability of exchange, the model or rule that transforms state into recommendation, the human or automated authority that may act, the uncertainty that must be shown, the provenance trail that must be retained, and the conditions under which the loop must stop. Such contracts are familiar in safety-critical engineering but are rarely explicit in biomedical informatics. Digital twins need them.

\section{A hierarchy of feedback loops}

Health twins are naturally hierarchical. At the smallest scale, a tissue or organ twin may couple imaging, electrophysiology, pathology, and molecular data with a model of local function. At the patient scale, a twin may connect clinical history, medications, behavior, wearable signals, imaging, and laboratory tests to treatment planning. At the enterprise scale, a twin may model the workflow through which care is delivered. At the consortium or population scale, a twin may represent distributed data generation, quality control, and knowledge translation. The Learning Health Systems promotes the principles of knowledge-generation 
as a part of the practice of medicine that leads to continual improvement in care~\cite{Friedman2010LHS,IOM2007LHS}.

The hierarchy is not merely a stack of data. Each level has a different feedback tempo and a different value proposition. An electrical signal in a neural circuit may require millisecond-level feedback. A surgical plan may evolve over days or weeks. A patient with dementia may be followed over years. A consortium producing reference brain-cell atlases may need feedback across tissue requests, sequencing, metadata capture, quality control, and data release. These loops can help one another, but only if their semantics are aligned.

This hierarchy explains why digital twins require several computing disciplines at once. Databases and knowledge graphs represent state and provenance. Formal methods specify transitions and invariants. Machine learning and simulation estimate hidden state and forecast futures. Human-computer interaction determines whether feedback is usable or harmful. Security and privacy govern who may observe or act. Distributed systems move data and models across institutions. Verification, validation, and uncertainty quantification decide when a recommendation is credible enough to influence the physical world.

Even multi-scale hierarchy may have its reductionistic limitations.
We explain everything only by breaking it into smaller parts: cells, molecules, atoms, and so on. 
The opposite view is pure holism, where we focus only on the whole and ignore the structure and autonomy of the parts, which is anti-digital twin. 
Perhaps  Koestler’s holarchy point is more suitable:
 in living and social systems, ``part'' and ``whole'' are relative terms: almost everything is both part and whole depending on the scale being considered
 and aspects of focus.

\section{Neuroinformatics is the stress test}

Neuroinformatics is a demanding testbed for digital twins because the nervous system resists single-scale representation. Brain health depends on molecular programs, cell types, circuits, vascular and metabolic state, behavior, environment, social support, medication, clinical workflow, and institutional capacity. Aging and dementia add long time horizons and shifting baselines. Epilepsy, Parkinson's disease, stroke, traumatic brain injury, psychiatric disorders, and neurodevelopmental conditions add heterogeneous phenotypes and interventions. No single model can be the twin.

The point, therefore, is not to build one giant brain simulation. The point is to build composable feedback loops that respect scale. A single-cell atlas can improve the interpretation of tissue vulnerability. Imaging and electrophysiology can localize circuit dysfunction. A patient trajectory model can connect seizures, cognition, medication, sleep, and adverse events. An enterprise workflow twin can ensure that the right imaging, care conference, surgical planning, follow-up, 
and documentation occur at the right time. 

Brain aging and dementia illustrate the need. A useful twin for Alzheimer's disease and related dementias would not only predict cognitive decline. It would influence which assessments are collected, when imaging or biomarker testing is indicated, how care partners are engaged, how trial eligibility is detected, how real-world outcomes are captured, and how population-level evidence returns to clinical decision support. The feedback loops include biology, behavior, care delivery, clinical trials, and evidence generation. Treating any one of these as the whole twin would be an oversimplification.

\section{Enterprise health twins}

Enterprise health twins make the feedback view concrete. Electronic health records achieved broad digitization, but they often function as administrative systems of record rather than cognitive systems of care~\cite{AdlerMilstein2017HITECH,Mandel2016SMART}. Clinicians face fragmented interfaces and documentation burdens that contribute to burnout~\cite{Shanafelt2016Burnout}. Specialty care is especially vulnerable because the clinically meaningful pathway often cuts across months, modalities, and teams.

The EpiToMe epilepsy-care system shows how a workflow can become a twin rather than a passive record. It is tethered to the enterprise EHR through health-data interfaces, grounded in the Epilepsy and Seizure Ontology~\cite{Sahoo2014EpSO}, and organized around the care pathway for complex epilepsy~\cite{Tao2021EpiToMe,Zhang2026EHT}. Its physical counterpart is not just an individual patient. It is the patient-care process: attending physicians, fellows, technologists, nurse navigators, conferences, diagnostic tests, reports, and follow-up.

The bidirectional loop is explicit. Feed-forward synchronization moves clinical events: admissions, reports, studies, imaging metadata, and documentation state, from the care environment into the digital counterpart. Feed-back support pushes context-aware checklists, status displays, missing-prerequisite alerts, and documentation assistance back into clinical work. The epilepsy enterprise health twin has supported seven production 
workflows and 119,485 clinical reports for 53,083 patients, with a 74.4\% reduction in case-conference documentation time and a nearly fourfold growth in completed surgical evaluations without added staffing~\cite{Zhang2026EHT}.

The important computing idea is not the particular dashboard. It is the representation of a clinical care pathway as a live state machine with
 typed events, ontology-grounded state, and guarded transitions. In EpiToMe, a patient journey tracker represents key steps in the epilepsy surgical evaluation pathway, and a surgical planning tool captures stereoelectroencephalography intent as structured data~\cite{Chou2025WISP}. This is feedback at enterprise scale: the digital counterpart changes what the care team sees, what it remembers, what it documents, what steps to follow, and what downstream research can reuse.

Enterprise feedback loops also show why the human must remain inside the twin. 
The virtual counterpart should not replace clinical judgment. It should reduce the clerical and coordination burden so judgment can focus on ambiguity, ethics, and patient goals. It should reveal missing evidence, 
but it should also show provenance, uncertainty, and the limits of automation. 

\section{Data generation is physical too}

An L4-level population health twin system is not common place. Here we use a prospective, large-scale national data generation ecosystem as an exemplar for this purpose.
A common mistake is to treat data production as outside the digital twin. In biomedical science, however, data generation is itself a physical process: specimens are requested, transported, dissected, assayed, sequenced, transformed, annotated, quality-controlled, archived, and reused. Every step can introduce delay, bias, loss, ambiguity, or irreproducibility. A digital twin of the data ecosystem can feed back into that process. 

The Tissue-to-Bytes effort in the BRAIN Initiative Cell Atlas Network is a useful example. It treats consortium-scale single-cell omics data production as a twinnable system: brain banks, laboratories, sequencing centers, archives, metadata standards, dashboards, common data elements, and provenance identifiers become part of a digital counterpart of the physical data-production enterprise~\cite{Tao2025TissueToBytes}. This shifts the twin concept from ``model of a brain'' to ``model of the system that generates evidence about brains.'' This consortium-scale digital twin is also multi-scale by definition: local institutional laboratory information management systems (LIMS) serves as feed-upward and feed-backward target systems (twined or not). 

Such a twin can change physical practice. It can identify missing metadata while specimens are still traceable. It can coordinate tissue requests across scarce donor material. It can expose quality-control failures before they propagate into public atlases. It can help decide which assays or brain regions should be prioritized. It can make provenance a living operational object rather than a retrospective publication supplement.

This matters for aging, dementia, and neurological disease research because real-world evidence is only as strong as the data-generating loops that produce it. A model trained on fragmented records, inconsistent phenotypes, or poorly tracked biospecimens may look sophisticated while amplifying invisible biases. 
Feedback-aware data generation turns infrastructure into science: the system learns not only from data, 
but also from the process by which the data came into being.

\section{Where the hard computing begins}

Making feedback first-class creates a research agenda that is broader than building better predictors. Five problems are especially urgent.

\textit{Specifying feedback contracts.} A digital twin should state what physical counterpart is twinned, what virtual state is maintained, what actions the virtual state can influence, what latency the loop assumes, and who is accountable for acting. Today, many systems describe models and data but leave the action channel implicit.

\textit{Composing hierarchical twins.} Organ, patient, health system, and population loops operate at different time scales and levels of abstraction. Composition requires typed interfaces, uncertainty propagation, and safeguards against false sense of precision. A single-cell finding should not automatically change a clinical plan; an enterprise workflow metric should not automatically define health outcome. The research problem is controlled translation across scales.

\textit{Validating actions, not only predictions.} Verification, validation, and uncertainty quantification are central to digital twins because the model is meant to act on the world~\cite{NASEM2024DigitalTwins,Laubenbacher2024Medicine}. Predictive accuracy is insufficient. We need to validate that the recommended action, display, alert, or workflow change improves the target physical system under realistic use.

\textit{Building interoperable provenance.} Health twins will fail if each is a bespoke island. Standards such as HL7 FHIR, domain ontologies, common data elements, FAIR principles, and provenance graphs must become operational infrastructure~\cite{Mandel2016SMART,Wilkinson2016FAIR}. 
The provenance of feedback matters: which data, model, version, rule, clinician, and institutional policy produced a recommendation?

\textit{Governing integration capacity.} A clinical team, laboratory network, or patient community can absorb only so much automated guidance before trust erodes or workarounds appear. Feedback loops therefore require governance, not only usability. Human-in-the-loop design should specify when humans initiate action, when they veto it, when automation is permitted, and how disagreements become data for improvement.

These problems are not peripheral to digital twins. They are the difference between a twin that learns responsibly and a dashboard that merely looks alive.

\section{A feedback-centered definition}

A digital twin is a governed computational counterpart of a specific physical system that is dynamically synchronized with that system, supports prediction or reasoning about possible states, and participates in bidirectional feedback that changes measurement, decisions, operations, or interventions in the physical world.

The definition excludes many useful artifacts. A data warehouse may be necessary, but it is not a twin. A visualization may be necessary, but it is not a twin. A machine-learning model may be necessary, but it is not a twin. A cyber-physical controller may be necessary, but it is not a twin unless it is embedded in a persistent, interpretable, updateable counterpart whose feedback can be audited and improved.

For neuroinformatics, this strict definition is liberating. It allows us to ask which feedback loops are worth building first. Organ-level twins can improve surgical and neuromodulation decisions. Patient-level twins can support longitudinal care and trial readiness. Enterprise twins can make EHR ecosystems learning-ready. Consortium twins can coordinate real-world data generation at scale. Together, they can advance brain health, aging, dementia, and neurological disease research without pretending that one model can contain the whole brain, the whole patient, and the whole health system.

The digital twin concept will earn its name only when bidirectional feedback becomes a first-class object. 
The future of the field is not a prettier mirror of the world. It is a disciplined way of closing loops between the world we care, the models we build, and the actions we take.

\begin{acks}
This work was supported in part by the National Science Foundation Award IIS2500624 and the National Institutes of Health grants U24MH130988,
 and U24AG098157. The views of the paper are those of the authors and do not reflect those of the funding agencies.
\end{acks}

\bibliographystyle{ACM-Reference-Format}

\end{document}